# A Novel Decoupled LVRT Control Strategy for Transient Voltage Stability Enhancement of IBRs Using Voltage-Angle Coupling Analysis

Fangyuan Sun[1], *Student Member*, *IEEE*, Ruisheng Diao[1*], *Senior Member*, *IEEE*, Ruiyuan Zeng[1], Jing Zhang[2], Jianguo Qian[2]

***Abstract*— With the fast-increasing penetration of inverter-based resources (IBRs), the voltage support capability of the grid following (GFL) IBRs under low voltage ride through (LVRT) control significantly influences the transient voltage stability of the power system. The existing LVRT adjusts the q-axis current to regulate reactive power injection; however, under a large disturbance, the phase-locked loop (PLL) error invalidates the proportional relationship between the q-axis current and reactive power, consequently causing deviation in the actual reactive power injection of the IBR. Besides, the variation of IBR's current, determined by the PLL phase and LVRT, also directly influences the transient voltage. To address this issue, the specific influence of PLL error on active and reactive power injection is first analyzed under LVRT control. In addition, by combining the LVRT and PLL dynamics, the mechanisms of three voltage problems caused by voltage angle coupling are revealed: overvoltage, low voltage, and DC-side overvoltage. The specific scenarios in which these voltage stability problems occur are also obtained by the voltage-vector-triangle graphic. Furthermore, a power angle decoupled LVRT control is proposed to eliminate the influence of voltage angle coupling. Finally, the mechanism analysis and effectiveness of the decoupled LVRT are verified in the case study.**

## I. INTRODUCTION

With the fast-increasing penetration of renewable energy, the dynamics of inverter-based resources become essential in affecting the transient stability of modern power systems. The GFL converter serves as the main interface for integrating renewable energy generators into power systems, and the low-voltage-ride-through (LVRT) control is widely adopted in short-circuit fault scenarios [1] for maintaining connectivity during fast transients. To provide reactive power support during low-voltage periods, the absolute value of the q-axis current is generally positively related to the voltage dip, allowing for the injection of reactive power. The LVRT plays a significant role in the transient dynamics of the GFL resources.

Phase-locked loop (PLL) is the most common synchronization unit adopted in the GFL resources [2],[3]. Plenty of research has indicated that under large disturbances, such as short-circuit faults or changes in grid impedance, there is a deviation between the phase of the PLL and the phase of the point of common coupling (PCC) voltage [4]. A certain period is required for the PLL to catch up with the PCC voltage phase. From the basic logic of the LVRT control, the adjustment of active and reactive power is achieved by adjusting the d-axis and q-axis currents. However, the d/q-axes are respectively proportional to active/reactive power only when the PLL phase aligns with the PCC voltage phase. This results in a deviation between the actual and ideal power outputs [5]. Considering that the time scale of the PLL dynamic is around 100ms, which is dramatically faster than the traditional synchronization dynamics but close to voltage stability dynamics [6], the active/reactive power fluctuation caused by the PLL dynamic can influence the transient voltage stability. This phenomenon is referred to as voltage angle coupling in this paper.

Many research efforts on the transient stability of the GFL devices considering LVRT and PLL dynamics have been reported. Current research can be divided into two categories: 1) stability analysis combining LVRT and PLL dynamics [7]-[11] and 2) new LVRT control based on transient dynamics [12]-[18].

For the stability analysis research, the PLL deviation under different grid strengths and grid fault is revealed in [7], and its influence on transient voltage support under different control strategies is also compared. In [8], the existence of the equilibrium points (EPs) of the GFL converter under LVRT control during grid fault is analyzed, and the non-existence of EP under certain system or control parameters is proved. In [9], the impact of LVRT control on the synchronization of the PLL is analyzed. Moreover, the estimation of the transient stability region under different LVRT parameters is given. In [10], the EPs of the GFL converter under LVRT are given from the voltage perspective, and the stability of the EPs is also derived to give a criterion of transient instability. In [11], the voltage-angle coupled transient stability is investigated using the voltage-vector-triangle graphic, and two types of instability shapes are identified. A corresponding criterion is also proposed for assessing the instability risk.

This work was supported by the State Grid Zhejiang Electric Power Company Research Project, "Research and Application of Key Technologies for the Construction of Artificial Intelligence Vertical Models for Power Grid Dispatching and Operation" (Project No. KJKY20240424).

[1]F. Sun, R. Diao, and R. Zeng are with the ZJU-UIUC Institute, Zhejiang University, Haining, 314400 China (e-mail: fysun_zju@zju.edu.cn, ruiyuan.23@intl.zju.edu.cn).

[2]J. Zhang and J. Qian are with State Grid Zhejiang Electric Power Company.

Corresponding author: Prof. Ruisheng Diao, email: ruishengdiao@intl.zju.edu.cn

For the research on the improvement of LVRT, a reference rematching method is proposed in [12] to adjust the d/q-axis current ratio in LVRT, and the existence of the EP is guaranteed during the fault period. In [13], the transient stability of the GFL resource is transformed into swing equation form, and an improved LVRT controlling the equivalent electrical power is proposed to avoid loss of synchronization. In [14], a frozen PLL during fault is proposed to guarantee synchronization stability during fault. The selection method for the LVRT parameter, the K-factor, is described in [15] to prevent loss of synchronism under a certain voltage drop. In [16], [17], and [18], apart from the converter dynamic, the dynamic of the power source, such as wind power system is also included. In [16], the transient process of the wind generation system is divided into several stages according to the switching control method. In [17], a communication-independent LVRT control for offshore wind farms is proposed for frequency support during the post-fault period. The fault-induced frequency deviation is reduced by adjusting the active power ramp rate. In [18], the transient stability of the DC link voltage of the doubly-fed induction generator (DFIG) is investigated by transforming the dynamic model to a swing equation form. An additional damping and slip power feedforward control strategy is proposed to enhance the stability of the DC voltage.

The current research gap lies in two aspects:

1) The current research are generally from the prespective of the converters, and the stability analysis mainly focus on synchronization stability. The DC voltage stability are studied in [17]-[18], respectively, however, the transient voltage stability at the AC side is insufficient. The common LVRT can provide voltage support in low-voltage period, but the PLL dynamics can support can deteriorate this supportive effect or even produce adverse impacts. This problem is primarily studied in some researches, but more profound analysis is required. In [11], it is illustrated that the angle dynamics of the GFL converter can influence the post-fault voltage magnitude, but the specific condition generating low-voltage and high-voltage tendency are not discussed. The influence on error of reactive power injection is not analyized either. In [5], [7], and [19], the voltage angle coupling is only mentioned, but it is not the core of the papers and thorough analysis is not given.

2) For the researches on improved LVRT or other control loops, current research efforts mainly focus on avoiding desynchronization of the device, but its stability support for the system has not received adequate attention. In [13], control method for frequency support is proposed, but transient voltage support is not included. How to eliminate the negative effect of voltage angle coupling requires further research.

To address the aforementioned issues, the voltage angle coupling, namely, the influence of PLL on transient voltage dynamics, is analyzed in this paper from two aspects: 1) the indirect influence due to introducing power injection error; and 2) the direct influence of changing initial post-voltage magnitude. Additionally, a power-angle decoupled control method is proposed to eliminate the influence of voltage-angle coupling. The main contributions of this paper are summarized as follows:

1) By combining LVRT control and PLL dynamics, the influence of PLL deviation on power injection is analyzed.

2) Based on the PLL dynamic, the mechanisms of the PLL leading/lagging/unchanged dynamic are revealed under different fault types.

3) Influences of various PLL dynamics on transient voltage are revealed from both aforementioned aspects. Furthermore, caused by voltage angle coupling, the relationships between various fault types and their corresponding voltage stability problems are established.

4) A power angle decoupled LVRT control is proposed to eliminate the influence of PLL dynamics on power injection. The existence of EP and transient stability domain are also analyzed.

The remainder of the paper is organized as follows. Section II introduces the model of the GFL resource-connected system. In Section III, the voltage angle coupling analysis is given in detail. In Section IV, the power angle decoupled LVRT strategy is proposed. Transient simulation results and the effectiveness verification of the power angle decoupled LVRT are given in Section V. Finally, Section VI draws the conclusion.

## II. System Modeling

A GFL resource-connected system is shown in Fig. 1, and the following assumptions are made. 1) the power system is represented as an ideal voltage source, whose voltage is given by $U_g\angle 0$. 2) A load model, including an induction motor(IM) and c onstant impedance load, is connected between the converter and the voltage source. The equivalent model of the induction motor is also shown in Fig. 1, which consists of stator impedance $\boldsymbol{Z_s}$, rotor reactance $X_r$, rotor resistance $R_r$, and exciting reactance $X_m$. 3) the GFL resource is represented as a controlled current source with a fixed current magnitude, $I_c$, and the phase angle $\delta_c$ is determined by the phase-locked loop (PLL). The phase angle between $\boldsymbol{I_c}$ and the d-axis of PLL is represented by $\theta_c$, and the phase angle between the PCC voltage $\boldsymbol{U_c}$ and the d-axis of PLL is represented by $\theta_v$. 4) The influence of frequency fluctuation is small in power systems, so its influence on reactance is ignored; that is, the reactances in Fig. 1 are constant. **Boldface** denotes complex variables, while non-bold variables represent either scalars or the magnitudes of complex variables.

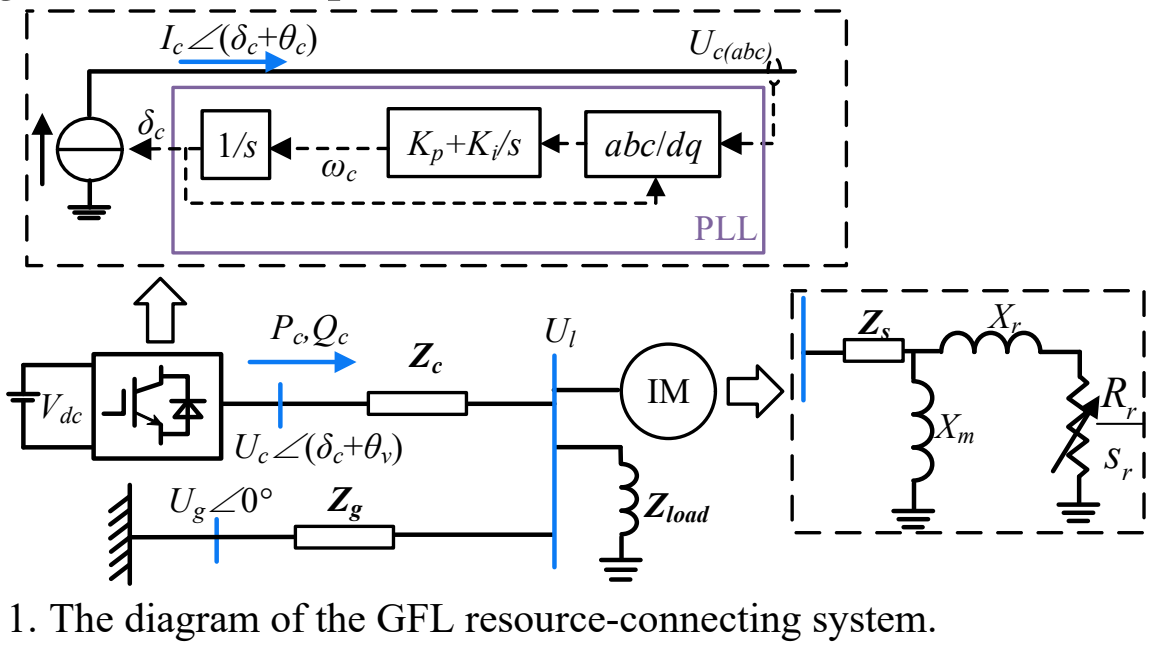

Fig. 1. The diagram of the GFL resource-connecting system.

Synchronization of the GFL resource is realized by PLL.

The diagram of the widely used synchronous reference frame PLL (SRF-PLL) is also shown in Fig. 1, whose dynamics can be expressed as:

$$\omega_c = K_p U_{cq} + \int K_i U_{cq} dt \\ \delta_c = \delta_{c0} + \int \omega_c dt \tag{1}$$

Where, $\omega_c$ is the angular velocity difference between the PLL phase and the fixed grid phase $\omega_0$ (100π or 120π rad/s). $K_i$ and $K_p$ are the integral and proportional coefficients of PLL. $U_{cq}$ is the q2-axis PCC voltage. $\delta_{c0}$ is the steady PLL phase angle.

LVRT control of GFL resources is typically required to inject reactive power under low-voltage conditions. In common LVRT control, the q-axis current $I_{cq}$ is proportional to the voltage deviation, while the remaining capacity is utilized to generate d-axis current $I_{cd}$, provided that the $I_{cd}$ does not exceed its initial reference value $I_{cd,ref}$. $I_{cq}$ and $I_{cd}$ can be expressed as:

$$I_{cq} = \begin{cases} -I_{\max}, U_c < U_{low} - \dfrac{I_{\max}}{K_q} \\ -K_q(U_{low} - U_c), U_c \geq U_{low} - \dfrac{I_{\max}}{K_q} \end{cases} \\ I_{cd} = \min\left(\sqrt{I_{\max} - I_{cq}^2}, I_{cd,ref}\right) \tag{2}$$

Where, $U_{low}$ is the LVRT activation voltage, and it is generally 0.9 p.u. $K_q$ is the proportion factor of reactive current injection. $I_{max}$ is the maximum current of the converter.

The first-order IM model is used to analyze the active and reactive power consumption during the transient process. The equivalent circuit is shown in Fig. 1, and $s_r$ is the slip ratio of the IM.

From (1) and (2), the variables $\omega_c$, $\delta_c$, $I_{cd}$, and $I_{cq}$ are decided by the PCC voltage $\boldsymbol{U}_c$, which can be calculated by the circuit equation of Fig. 1:

$$U_c e^{j(\delta_c + \theta_v)} = \boldsymbol{Z}_{eq} \boldsymbol{I}_c + \boldsymbol{U}'_g \\ \boldsymbol{I}_c = \left(I_{cd} + jI_{cq}\right) e^{j\delta_c} = I_{\max} e^{j(\delta_c + \theta_c)} \\ \boldsymbol{Z}_{eq} = \boldsymbol{Z}_g * \boldsymbol{Z}_l / \left(\boldsymbol{Z}_g + \boldsymbol{Z}_l\right) + \boldsymbol{Z}_c \\ \boldsymbol{U}'_g = \boldsymbol{U}_g * \boldsymbol{Z}_l / \left(\boldsymbol{Z}_g + \boldsymbol{Z}_l\right) \tag{3}$$

Where, $\boldsymbol{U}'_g$ and $\boldsymbol{Z}_{eq}$ represent the Thevenin-equivalent voltage and impedance of the circuit. $\boldsymbol{Z}_l$ is the equivalent impedance of the load. Combining (2) and (3), $I_{cd}$, $I_{cq}$, $U_c$, and $\theta_v$ can be solved under a certain $\delta_c$. It is worth noting that the combined equations may have no solution, leading to a loss of synchronization. This problem has been studied in [8], and when the parameter $K_q$ is properly set, instability may occur under large power angle deviation. So, this condition will not be discussed in this paper.

## III. Voltage-Angle Coupling Analysis

### A. Power Injection Error Due to PLL Error

The purpose of LVRT control is to prioritize reactive power injection over active power as voltage sags deepen, allocating more converter capacity to VAR support, which is realized by proportional adjusting q-axis current $I_{cq}$. This is only tenable under steady state, when the PLL phase aligns with the PCC voltage phase, which is shown in Fig. 2 (a). The active and reactive power of the converter can be expressed as:

$$P_c = I_{cd} U_c, Q_c = I_{cq} U_c \tag{4}$$

Where, $P_c$ and $Q_c$ are the active and reactive power outputs of the converter.

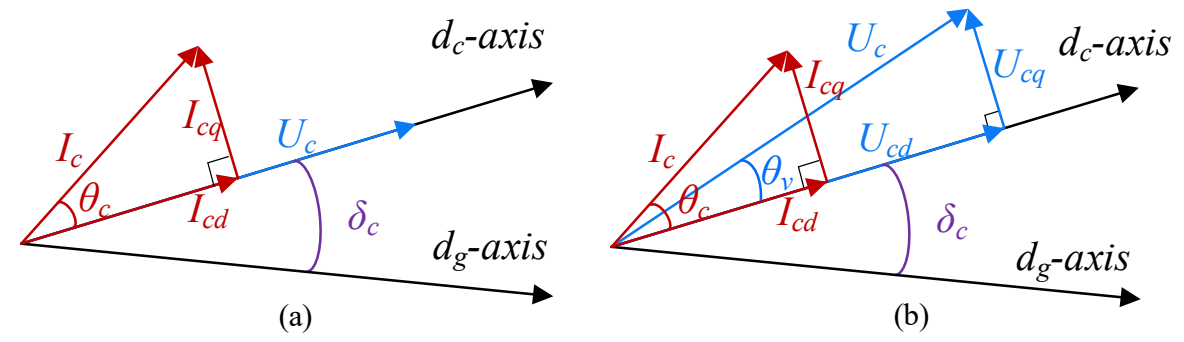

Fig. 2. Phase diagram of GFL resource under (a) steady state and (b) transient state.

However, LVRT is likely to be activated under large disturbances, such as short-circuit faults or other phase jump conditions. During the transient process of large disturbances, there will be a deviation between the PLL phase and the phase of $\boldsymbol{U}_c$, which is $\theta_v$ shown in Fig. 2 (b). The active and reactive power of the converter can be expressed as:

$$Q_c = I_{cd} U_{cq} - I_{cq} U_{cd} = I_{cd} U_c \sin\theta_v - I_{cq} U_c \cos\theta_v \\ P_c = I_{cd} U_{cd} + I_{cq} U_{cq} = I_{cd} U_c \cos\theta_v + I_{cq} U_c \sin\theta_v \tag{5}$$

From (5), the PLL deviation $\theta_v$ makes $Q_c$ no longer proportional to $I_{cq}$, but also influenced by $I_{cd}$ and $\theta_v$. This means that a lower $U_c$ does not always correspond to a higher reactive power injection $Q_c$, resulting in a non-ideal $Q_c/P_c$ injection. By combining (2) with (4) and (5) respectively, the relationship between $Q_c/P_c$ and $U_c/\theta_v$ under LVRT control is shown in Fig. 3, with the parameters $I_{max}$=1.2 p.u., $U_{low}$=0.9 p.u., and $K_q$=2. The blue surface corresponds to (5) and Fig. 2 (b), when PLL deviation is considered, and the orange surface corresponds to (4) and Fig. 2 (a). The $\theta_v$ isn't included in (4), and $Q_c/P_c$ does not change in the $\theta_v$ dimension. The following conclusions can be drawn from Fig. 3:

1) From Fig. 3 (a), when the PLL phase lags $\boldsymbol{U}_c$ phase ($\theta_v$>0), there exists a voltage threshold such that when $U_c$ exceeds this value, the actual $Q_c$ surpasses the ideal $Q_c$. This voltage threshold increases with $\theta_v$. When the PLL phase leads $\boldsymbol{U}_c$ phase ($\theta_v$<0), the actual $Q_c$ is always less than the ideal $Q_c$. Moreover, negative $Q_c$ tends to occur when the absolute value of $\theta_v$ is large, especially when $\theta_v$ is negative.

2) From Fig. 3 (b), when the PLL phase lags $\boldsymbol{U}_c$ phase ($\theta_v$>0), the actual $P_c$ is always less than the ideal $P_c$. When the PLL phase leads $\boldsymbol{U}_c$ phase ($\theta_v$<0), There exists a voltage threshold such that when $U_c$ exceeds this value, the actual $P_c$ is less than the ideal $P_c$. This voltage threshold increases with $\theta_v$. Moreover, negative $P_c$ tends to occur when the absolute value of $\theta_v$ is large, especially when $\theta_v$ is positive.

The $Q_c/P_c$ injection deviation caused by PLL synchronization error is shown in Fig. 3. This synchronization problem will further result in voltage issues: 1) Insufficient $Q_c$ and excessive $P_c$ may lead to low voltage problems, such as slow voltage recovery, converter tripping, or IM stalling; 2) Excessive $Q_c$ and insufficient $P_c$ may cause overvoltage problems; and 3) Low or negative $P_c$ may also induce DC-side overvoltage problems.

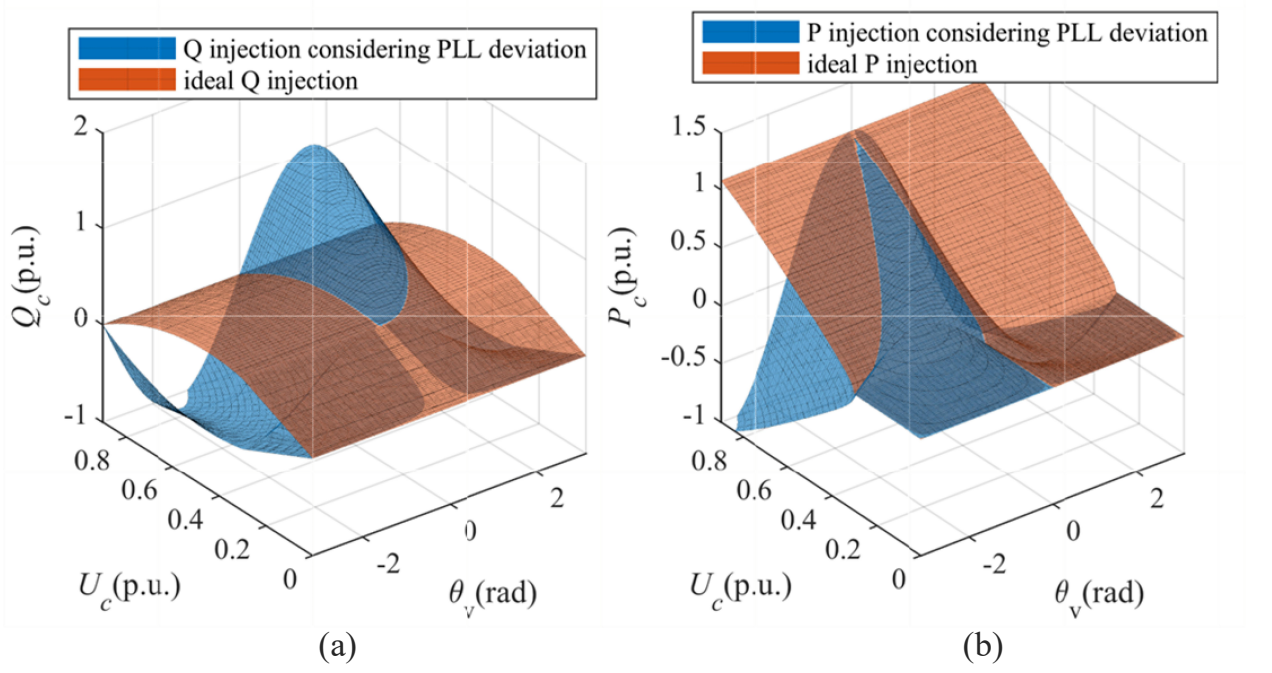

Fig. 3. The relationship between $Q_c$/$P_c$ and $U_c$/$\theta_v$ under LVRT control.

The analysis of these voltage stability problems inherently involves the PLL-dominated power angle dynamics:

When a fault occurs, there is a mutation of PCC voltage $\boldsymbol{U_c}$, both in magnitude and phase angle, and $U_{cq}$ is detected. From (1), the angular velocity difference $\omega_c$ will accumulate until $U_{cq}$ is zero, and hence, phase angle $\delta_c$ will also change. After the fault is cleared, there will be a deviation between the current $\delta_c$ and its pre-fault value. The changed $\delta_c$ and converter current $\boldsymbol{I_c}$, alongside the post-fault circuit, decide the magnitude and phase of $U_c$ by the power flow equation (3), and the $U_c$ will, in turn, influence $\boldsymbol{I_c}$ through LVRT control (2). By combining (2), (3), and (5), it can be concluded that with a certain post-fault $\delta_c$, which is decided by the on-fault transient, $\boldsymbol{U_c}$ and $\boldsymbol{I_c}$ can be decided, and then, $Q_c$ and $P_c$.

### B. *PLL Error Analysis and the Influence on Voltage Dynamics*

From the aforementioned analysis, $\delta_c$ decides $Q_c$/$P_c$ injection and further decides which kind of PLL error-related voltage problem may occur. Besides, the change of the PLL phase during the fault period, along with LVRT control, can influence the initial post-fault voltage through the phase variation of $\boldsymbol{I_c}$, and further influence the post-fault voltage recovery. These are the two mechanisms through which the angle dynamics of PLL influence transient voltage.

Two transient periods, on-fault and post-fault, are analyzed separately in this section. The dynamics of power angle $\delta_c$ are analyzed during the on-fault period, and in the post-fault period, we focus on how the PLL resynchronization process impacts voltage stability.

The on-fault system circuit can be approximated, as shown in Fig. 4 (a), where the voltage source side of the circuit is bypassed by the short-circuit impedance. $\boldsymbol{Z_{cf}}$ is the impedance between the PCC point and the fault point, and $\boldsymbol{Z_f}$ is the fault impedance. Three cases are considered considering different fault locations: PCC fault, nearby fault, and remote fault.

1) For PCC fault, $\boldsymbol{Z_{cf}}$ is zero, and $\boldsymbol{Z_f}$ is close to zero. It is foreseeable that the PCC voltage $U_c$ is also close to zero. From the LVRT strategy in (2), $I_{cq}$=-$I_{max}$ and $I_{cd}$=0, and $\boldsymbol{I_c}$ lags the PLL phase by 90°. The phase diagram is shown in Fig. 4 (b). Since $\boldsymbol{U_c}$= $\boldsymbol{I_c}$*$\boldsymbol{Z_f}$ is close to zero, its q-axis component $U_{cq}$ is also close to zero. From (1), though the PLL phase is not aligned to $\boldsymbol{U_c}$, $\delta_c$ will be changed very slowly due to the small $U_{cq}$. So, under the PCC fault scenario, $\delta_c$ is hardly changed during the on-fault period.

2) For nearby faults, $\boldsymbol{Z_{cf}}$ is small but not close to zero, which makes $\boldsymbol{Z_f}$ negligible. The PCC voltage $U_c$ will also be less than $U_{low}$-$I_{max}$/$K_q$. From (2), $I_{cq}$= − $I_{max}$ and $I_{cd}$=0, and $\boldsymbol{I_c}$ lags the PLL phase by 90°. The phase diagram is shown in Fig. 4(c). Using $X_{cf}$ and $R_{cf}$ to represent the reactance and resistance of $Z_{cf}$, if $R_{cf}$ is negligible, $\boldsymbol{U_c}$= $\boldsymbol{I_c}$*j$X_{cf}$ will lead $\boldsymbol{I_c}$ by 90°, and will be exactly aligned to the PLL phase. This means $U_{cq}$=0, and $\delta_c$ is not changed during the on-fault period. When $R_{cf}$ is not negligible, from the dashed vector $U_c$ in Fig. 4 (c), $\boldsymbol{U_c}$ always lags the PLL phase, and $U_{cq}$ is a fixed negative value. From (1), $\delta_c$ will keep reducing during the on-fault period with an accelerating speed.

3) For remote faults, $\boldsymbol{Z_{cf}}$ is larger, which makes the PCC voltage $U_c$ larger than $U_{low}$-$I_{max}$/$K_q$. From (2), $I_{cq}$∈[ − $I_{max}$, 0], and $I_{cd}$ ∈[0, $I_{cd,ref}$]. $\boldsymbol{I_c}$ lags the PLL phase with a phase angle less than 90°. The phase diagram is shown in Fig. 4 (d). When $R_{cf}$ is negligible, $\boldsymbol{U_c}$= $\boldsymbol{I_c}$*j$X_{cf}$ leads $\boldsymbol{I_c}$ by 90°, and also leads the PLL phase. $U_{cq}$ is a fixed positive value. From (1), $\delta_c$ will keep increasing during the on-fault period with an accelerating speed. When $R_{cf}$ is not negligible, the phase angle between $\boldsymbol{U_c}$ and $\boldsymbol{I_c}$ reduces. $U_{cq}$ is likely to be a fixed positive value, and $\delta_c$ will keep increasing during the on-fault period with an accelerating speed. But when $U_c$ is close to $U_{low}$-$I_{max}$/$K_q$ or $R_{cf}$ is large, $U_{cq}$ can be a fixed negative value, and $\delta_c$ will keep reducing during the on-fault period with an accelerating speed.

After fault clearance, the PLL phase will resynchronize to the original position, during which the change of $\delta_c$ and post-fault circuit can influence the PCC voltage and the voltage recovering process. To analyze this influence, the post-fault circuit is firstly given in Fig. 5. The load impedance $\boldsymbol{Z_l}$ is composed of impedance load and IM load. Through Thevenin's theorem, the post-fault circuit can be equivalently represented as an ideal voltage source $\boldsymbol{U_g'}$ in series with an impedance $\boldsymbol{Z_{eq}}$, which is also shown in Fig. 5. $\delta_g$ and $\theta_z$ are the phase angles of $\boldsymbol{U_g'}$ and $\boldsymbol{Z_{eq}}$. In this analysis, $\delta_c$ is limited in the range [-80°, 130°], because excessive $\delta_c$ deviation may lead to loss of synchronization of the PLL, which is not the focus of this paper.

The phase diagram of the post-fault system in Fig. 5 is shown in Fig. 6, with the parameters: $I_{max}$=1.2 p.u., $U_{low}$=0.9 p.u., $K_q$=1.5, $\boldsymbol{U_g}$=1.1 p.u., $\boldsymbol{Z_g}$=j0.2 p.u., $\boldsymbol{Z_c}$=j0.2 p.u., $\boldsymbol{Z_l}$=(0.495+j0.0495) p.u. The load is 100% impedance load, and from Thevenin's theorem, $\boldsymbol{U_g'}$ =(0.9214-j0.3544) p.u., $\boldsymbol{Z_{eq}}$=(0.0644+j0.3675) p.u.

From the on-fault circuit in Fig. 5, $U_c$=|$\boldsymbol{I_c}$* $\boldsymbol{Z_c}$|>$U_{low}$-$I_{max}$/$K_q$, which means the fault can be categorized as a remote fault, and $\delta_c$ will keep increasing during the on-fault period. In Fig. 6, the fault is cleared when $\delta_c$ increases to 130°. The pre-fault vector of $\boldsymbol{U_c}$ and $\boldsymbol{I_c}$*$\boldsymbol{Z_{eq}}$ are represented by the dashed vectors in Fig. 6. Comparing to the pre-fault vector $\boldsymbol{I_c}$*$\boldsymbol{Z_{eq}}$, the increase of $\delta_c$ leads $\boldsymbol{I_c}$ and $\boldsymbol{I_c}$*$\boldsymbol{Z_{eq}}$ to rotate counterclockwise. The purple and blue dashed circles represent the length of pre-fault $\boldsymbol{U_c}$ and post-fault $\boldsymbol{I_c}$*$\boldsymbol{Z_{eq}}$. With the change of the $\boldsymbol{I_c}$ phase

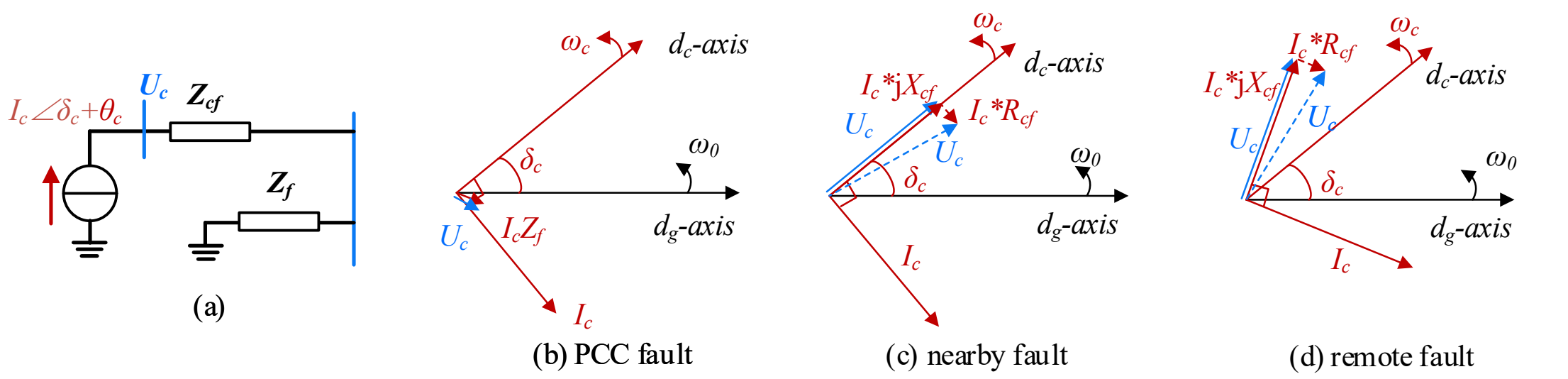

Fig. 4. (a) On-fault equivalent circuit, (b) phase diagram under PCC fault, (c) phase diagram under nearby fault, (d) phase diagram under remote fault.

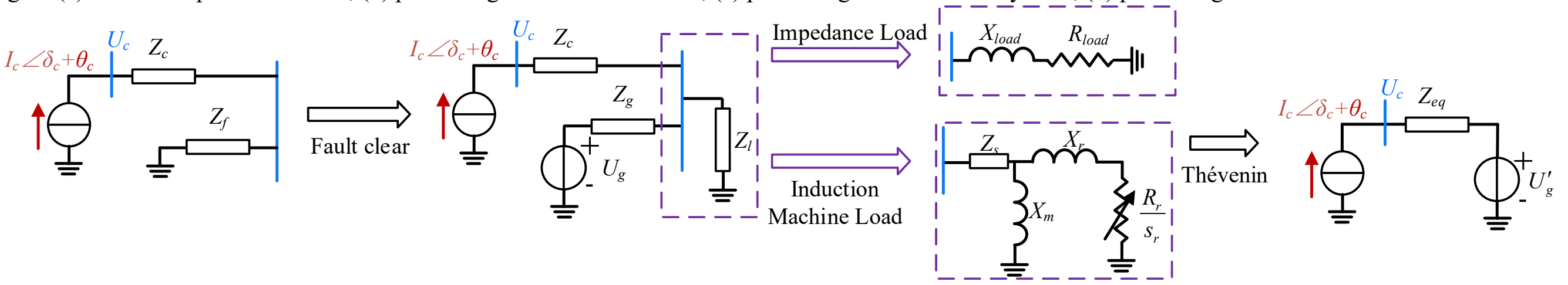

Fig. 5. Post-fault equivalent circuit.

($\delta_c$ and $\theta_c$), the vector $\boldsymbol{I_c}$*$\boldsymbol{Z_{eq}}$ will rotate inside the blue dashed circle, and $\boldsymbol{U_c}$ will also change accordingly. From the geometrical relationship, when $U_c$ equals its pre-fault $U_{c0}$ value, the vector relationships are as follows:

$$\begin{cases} U_{c0} = U_c \\ \boldsymbol{U_{c0}} = \boldsymbol{U'_g} + \boldsymbol{I_{c0}}\boldsymbol{Z_{eq}} = \boldsymbol{U'_g} + I_{c0}Z_{eq}e^{j(\delta_{c0}+\theta_{c0}+\theta_z)} \\ \boldsymbol{U_c} = \boldsymbol{U'_g} + \boldsymbol{I_c}\boldsymbol{Z_{eq}} = \boldsymbol{U'_g} + I_cZ_{eq}e^{j(\delta_{c,cr}+\theta_{c,cr}+\theta_z)} \\ (4) \end{cases} \quad (6)$$

Where, $\boldsymbol{U_{c0}}$, $\boldsymbol{I_{c0}}$, $\delta_{c0}$, and $\theta_{c0}$ are the pre-fault $\boldsymbol{U_c}$, $\boldsymbol{I_c}$, $\delta_c$, and $\theta_c$. $\delta_{c,cr}$ and $\theta_{c,cr}$ are the critical values of $\delta_c$, and $\theta_c$, satisfying $U_c$=$U_{c0}$. The equation (3) is used to calculate $\theta_{c,cr}$.

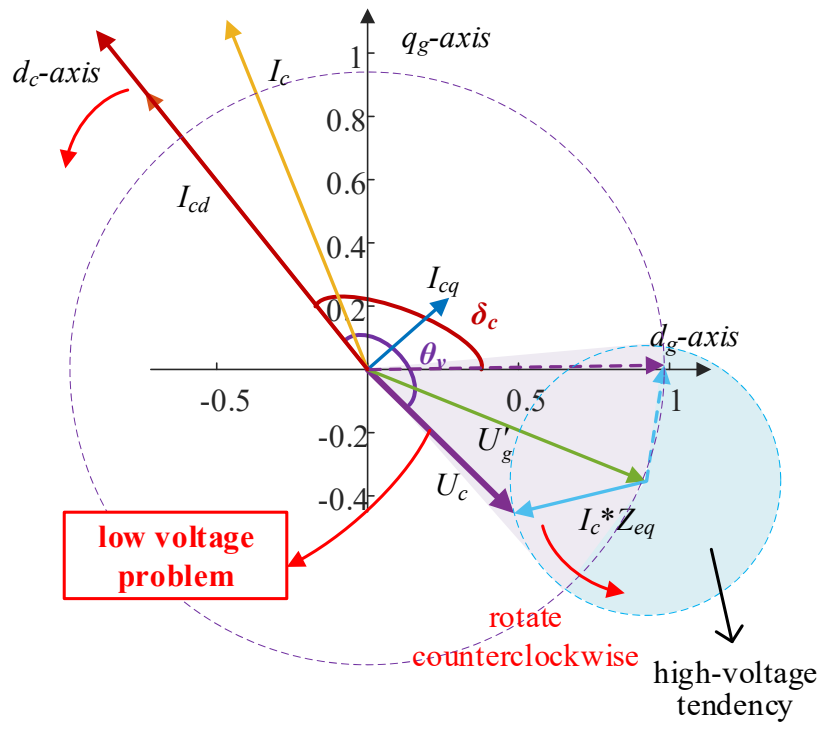

Fig. 6. Phase diagram after clearance of remote fault.

From (6), the blue area inside the blue dashed circle represents the direction of $\boldsymbol{I_c}$*$\boldsymbol{Z_{eq}}$ that results in a larger $\boldsymbol{U_c}$ magnitude than the pre-fault value. On the contrary, the counterclockwise rotation caused by the increasing $\delta_c$ reduces the $\boldsymbol{U_c}$ magnitude, which means the PCC voltage will recover from a low value after fault clearance due to the PLL dynamics. It can also be concluded from the geometrical relationship that the direction of $\boldsymbol{U_c}$ only locates inside the purple area, which most likely lags the PLL phase, leading to a negative $\theta_v$. From Fig. 3, the reactive power injection is less than ideal value. In summary, under remote fault, the increasing $\delta_c$ generally results in a lower initial value of the post-fault recovery. The PLL phase will also lead the $\boldsymbol{U_c}$ phase ($\theta_v$ is negative), causing insufficient reactive power injection and exacerbating the low-voltage issue.

By changing the parameters $K_q$=2, $\boldsymbol{Z_c}$=(0.0684+j0.1879) p.u., the on-fault voltage $U_c$=|$\boldsymbol{I_c}$*$\boldsymbol{Z_c}$|<$U_{low}$-$I_{max}$/$K_q$, which means the fault can be categorized as a nearby fault, and $\delta_c$ will keep reducing during the on-fault period. The post-fault phase diagram under nearby fault is shown in Fig. 7 (a). The fault is cleared when $\delta_c$ decreases to -80°. The pre-fault vectors of $\boldsymbol{U_c}$ and $\boldsymbol{I_c}$*$\boldsymbol{Z_{eq}}$ are represented by the dashed vectors. With the decreasing of $\delta_c$, the vectors $\boldsymbol{I_c}$ and $\boldsymbol{I_c}$*$\boldsymbol{Z_{eq}}$ also rotate clockwise. The purple and blue dashed circles represent the length of pre-fault $\boldsymbol{U_c}$ and post-fault $\boldsymbol{I_c}$*$\boldsymbol{Z_{eq}}$. From (6), the blue area inside the blue circle corresponds to the direction of $\boldsymbol{I_c}$*$\boldsymbol{Z_{eq}}$, generating a larger $\boldsymbol{U_c}$ magnitude than the pre-fault value. The clockwise rotation exactly makes the vector of $\boldsymbol{I_c}$*$\boldsymbol{Z_{eq}}$ in this area, resulting in a larger $\boldsymbol{U_c}$ than its pre-fault value. Then, the converter will exit the LVRT control, and $I_{cq}$ reduces from a positive value to zero and the new phase diagram is shown in Fig. 7 (b). This further causes the vectors $\boldsymbol{I_c}$ and $\boldsymbol{I_c}$*$\boldsymbol{Z_{eq}}$ to rotate clockwise, keeping $\boldsymbol{U_c}$ above its pre-fault value when $\delta_c$ does not lag excessively. It can also be observed that the direction of $\boldsymbol{U_c}$ only locates inside the purple area. From the geometrical relationship, the phase of $\boldsymbol{I_c}$ always lags $\boldsymbol{U_c}$ when the PLL phase exceeds the purple area, which means $\theta_v$ is positive. When the PLL phase is located in the purple area, the direction of vector $\boldsymbol{I_c}$*$\boldsymbol{Z_{eq}}$ will most likely maintain a phase lead of $\boldsymbol{U_c}$ over $\boldsymbol{I_c}$. From Fig. 3, the positive $\theta_v$ and high $U_c$ cause a higher reactive power injection than the ideal value. Besides, when $\delta_c$ lags further excessively, larger positive $\theta_v$ can lead to negative active power from Fig. 3 (b), which will lead to an overvoltage problem at the DC side. In summary, under nearby faults, the reducing $\delta_c$ generally results in a higher initial value of the post-fault recovery. The PLL phase is likely to lag $\boldsymbol{U_c}$ phase ($\theta_v$ is positive), causing excessive reactive power injection and exacerbating the high-voltage issue. When the PLL phase lags $\boldsymbol{U_c}$ phase further excessively, overvoltage problems at the DC side may also occur due to the negative active power output.

TABLE I.
SUMMARY OF VOLTAGE-ANGLE COUPLING DYNAMIC

| **Fault type** | **Change of $\delta_c$** | **$U_c$ after fault clearance** | **Power injection error** | **Influence on voltage dynamic** |
|---|---|---|---|---|
| PCC fault | Negligible | Negligible influence | Negligible | Negligible |
| Nearby fault | Decrease ($R_{cf}$>0) | Higher $U_c$ (IM dynamic does not dominate or $\delta_c$>$\delta_{c,cr}$) | Higher $Q_c$ Lower $P_c$ | High voltage tendency |
| | | Lower $U_c$ (IM dynamic dominates and $\delta_c$<$\delta_{c,cr}$) | Lower $Q_c$ Lower $P_c$ | Low voltage tendency and high DC voltage tendency |
| | Negligible ($R_{cf}$=0) | Negligible influence | Negligible | Negligible |
| Remote fault | Increase | Lower $U_c$ | Lower $Q_c$ Higher $P_c$ (small $\delta_c$) | Low voltage tendency |
| | | | Lower $Q_c$ Lower $P_c$ (large $\delta_c$) | Low voltage tendency and high DC voltage tendency |

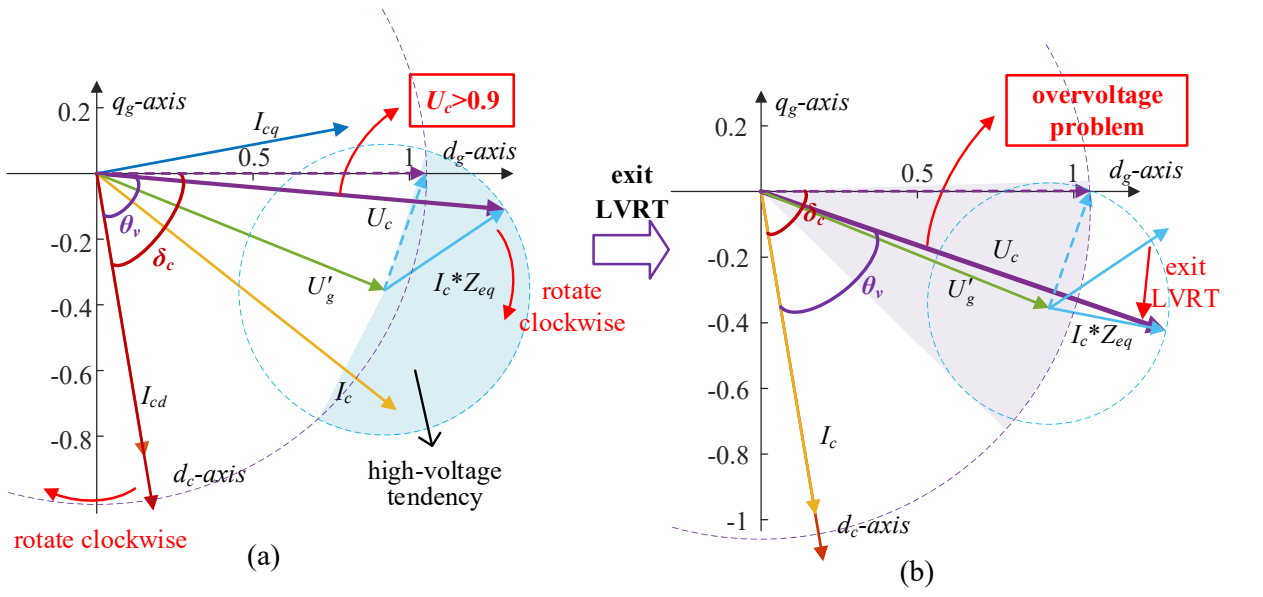

Fig. 7. Phase diagram after after clearance nearby fault.

When the dynamics of IM are considered, there will be a significant reduction of $\boldsymbol{Z_l}$ after fault clearance due to the increase of slip ratio $s_r$. Both $\boldsymbol{Z_{eq}}$ and $\boldsymbol{U'_g}$ will also reduce according to Thevenin's theorem, and the converter primarily faces low-voltage problems due to the dominance of this characteristic. Under this condition, the post-fault voltage is inevitably lower than the pre-fault voltage, making comparisons between them, like Fig. 6 or Fig. 7, meaningless. However, comparing the post-fault voltages under PLL phases $\delta_c$ and $\delta_{c0}$ allows analysis of whether PLL phase variation deteriorates the IM-dominated low-voltage problem. Taking the post-fault voltage under PLL phase $\delta_{c0}$ as the criteria $U'_{c0}$, the critical $\delta_c$ making $U_c$=$U'_{c0}$ can be calculated by:

$$\begin{cases} U'_{c0} = U_c \\ \boldsymbol{U'_{c0}} = \boldsymbol{U'_g} + \boldsymbol{I_{c0}}\boldsymbol{Z_{eq}} = \boldsymbol{U'_g} + I_{c0}Z_{eq}e^{j(\delta_{c0}+\theta'_{c0}+\theta_z)} \\ \boldsymbol{U_c} = \boldsymbol{U'_g} + \boldsymbol{I_c}\boldsymbol{Z_{eq}} = \boldsymbol{U'_g} + I_c Z_{eq}e^{j(\delta_{c,cr}+\theta_{c,cr}+\theta_z)} \\ (4) \end{cases} \quad (7)$$

There are two differences between (6) and (7). Firstly, the pre-fault voltage $U_{c0}$ is replaced by $U'_{c0}$ (post-fault voltage when $\delta_c$=$\delta_{c0}$). Secondly, in (6), $\theta_{c0}$ is a fixed parameter decided by the rated active and reactive power output, but in (7), $\theta'_{c0}$ is also calculated from (3), like $\theta_c$.

The phase diagram after fault clearance with IM load is shown in Fig. 8. From (7), the blue area inside the blue dashed circle represents the direction of $\boldsymbol{I_c}$*$\boldsymbol{Z_{eq}}$ with high post-fault voltage tendency, which mitigates the low-voltage problem caused by IM load, while the white area indicates the low-voltage problem deteriorates. In Fig. 8, $\delta_c$ is exactly the $\delta_{c,cr}$ calculated in (7), which is -156.01°. Compared to Fig. 6 and Fig. 7, the high-voltage tendency area in Fig. 8 is narrowed. This is caused by the low-voltage condition and LVRT control in (2). As a result, when the lag of PLL phase is large, a low-voltage tendency may arise, which contradicts the results in Fig. 7. Besides, from Fig. 3 (a), the low-voltage condition also narrows the range of $\theta_v$ that injecting excessive reactive power, also deteriorates the low-voltage problem. In summary, the dynamic of IM load not only generates a low-voltage problem but also makes the voltage angle coupling more prone to have a low-voltage tendency, further deteriorating the low-voltage problem. More specifically, there is a high-voltage tendency when the lag of the PLL phase is relatively small, and a low-voltage tendency occurs when the PLL phase exhibits lead or large lag.

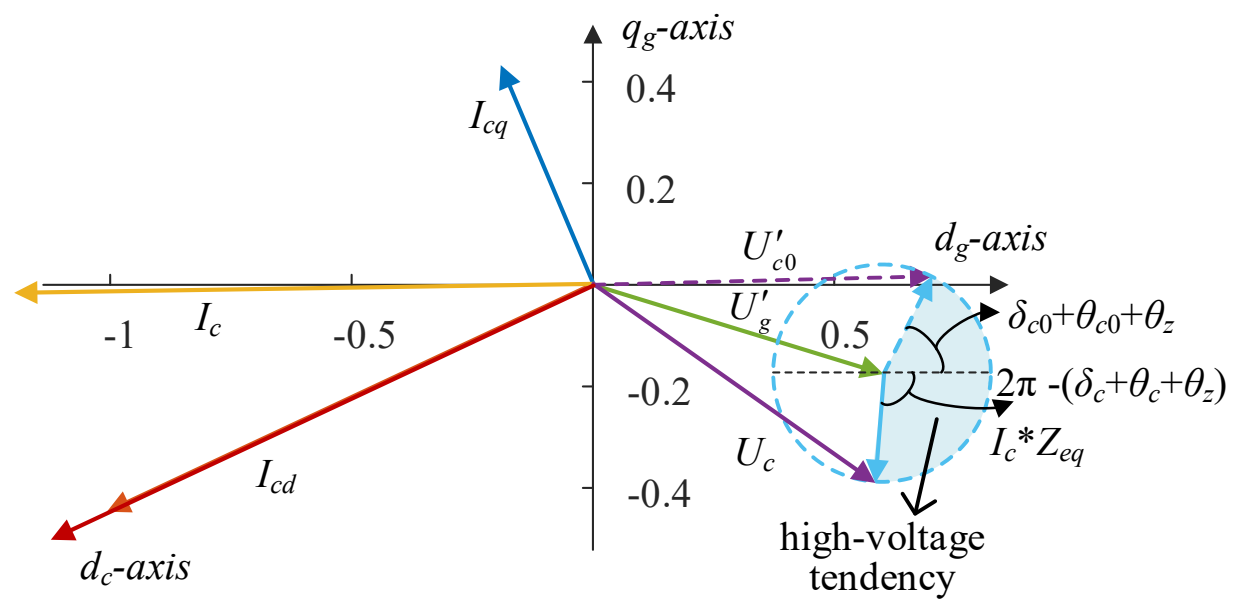

Fig. 8. Phase diagram after fault clearance considering IM load.

The above conclusions are drawn under specific parameters. However, in general systems, the different parameters mainly influence the size of the blue circle. This will cause deviations in the conclusions of Fig. 6, Fig. 7, and Fig. 8 when $\delta_c$ exhibits minor deviations, but the validity of these conclusions remains unaffected under larger $\delta_c$ deviations, which is much more critical.

In Table I, the voltage angle coupling dynamic is summarized, and under four cases, the PLL dynamics influence post-fault voltage. 1) Under nearby fault, if $R_{cf}$ is not negligible, $\delta_c$ will decrease, and the PLL dynamic causes high voltage tendency. 2) Under the same condition in 1), if the IM dynamic dominates and $\delta_c$ lagging is large, the PLL dynamics will cause a low voltage tendency. 3) Under remote fault, $\delta_c$ will increase, and the PLL dynamic causes low voltage tendency. 4) Under remote fault, when $\delta_c$ is large, causing a negative $\theta_v$ with a large absolute value, PLL error causes high voltage tendency at the DC side.

## IV. POWER-ANGLE DECOUPLED LVRT CONTROL

### A. Decoupled LVRT control strategy

From the analysis in Section III, the deviation between the actual and ideal power injection is caused by the deviation between the PCC voltage phase and the PLL phase. However, since the magnitude $U_c$ and q-axis component $U_{cq}$ are already used in LVRT control and PLL, respectively, the phase deviation $\theta_v$ is easily obtained using inverse trigonometric functions. On the basis of the traditional LVRT, a phase compensation of $\theta_v$ can be added to the $\boldsymbol{I_c}$ phase. The phase diagram is shown in Fig. 9, and the calculation of $I_{cd}$ and $I_{cq}$ can be expressed as:

$$\begin{aligned} I_{cq} &= I'_{cq}\cos\theta_v + I'_{cd}\sin\theta_v \\ I_{cd} &= I'_{cd}\cos\theta_v - I'_{cq}\sin\theta_v \end{aligned} \tag{8}$$

$$I'_{cq} = \begin{cases} -I_{\max}, U_c < \dfrac{I_{\max}}{K_q} - U_{low} \\ -K_q(U_{low} - U_c), U_c \ge \dfrac{I_{\max}}{K_q} - U_{low} \end{cases} \tag{9}$$

$$I'_{cd} = \min\left(\sqrt{I_{\max} - I'^2_{cq}}, I_{cd,ref}\right)$$

Where, $I'_{cd}$ and $I'_{cq}$ are the projections of $\boldsymbol{I_c}$ in the directions parallel and perpendicular to $\boldsymbol{U_c}$, respectively. From Fig. 9, it can be seen that by introducing the phase compensation of $\theta_v$, the $\boldsymbol{I_c}$ phase is decoupled with the PLL phase. It can also be concluded that:

$$P_c = I'_{cd}U_c, Q_c = I'_{cq}U_c \tag{10}$$

Comparing $\begin{cases}(2)\\(4)\end{cases}$ and $\begin{cases}(9)\\(11)\end{cases}$, the power injection $Q_c$ and $P_c$ of decoupled control are equal to that of ideal power injection, and the voltage problems caused by PLL error can be avoided.

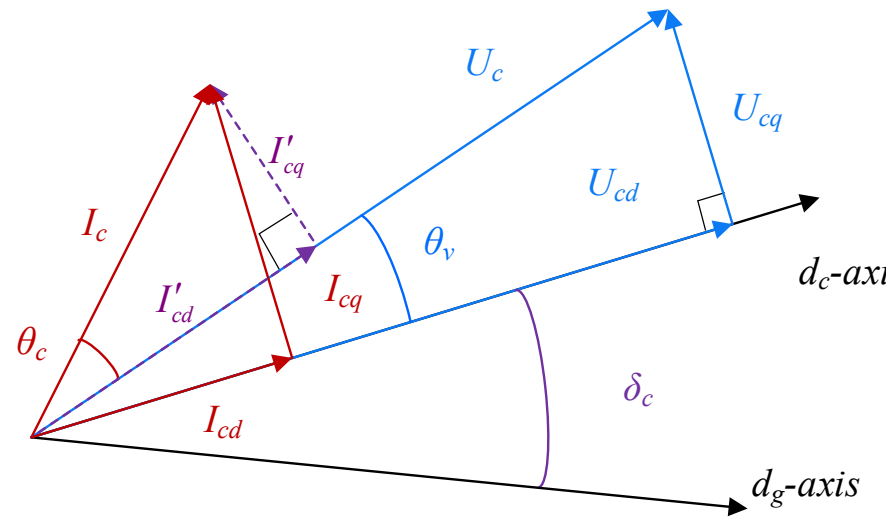

Fig. 9. The phase diagram of the decoupled LVRT control

## B. Stability analysis of decoupled control

The diagram of LVRT control is shown in Fig. 10, and the decoupled control part is highlighted in blue. In such a novel design, $\theta_v$ is not only a power flow result but also influences $I_c$, which in return influences the power flow. This feedback loop may lead to the inexistence of power flow solution, i.e. EP when combing the equations (8), (9) and (3).

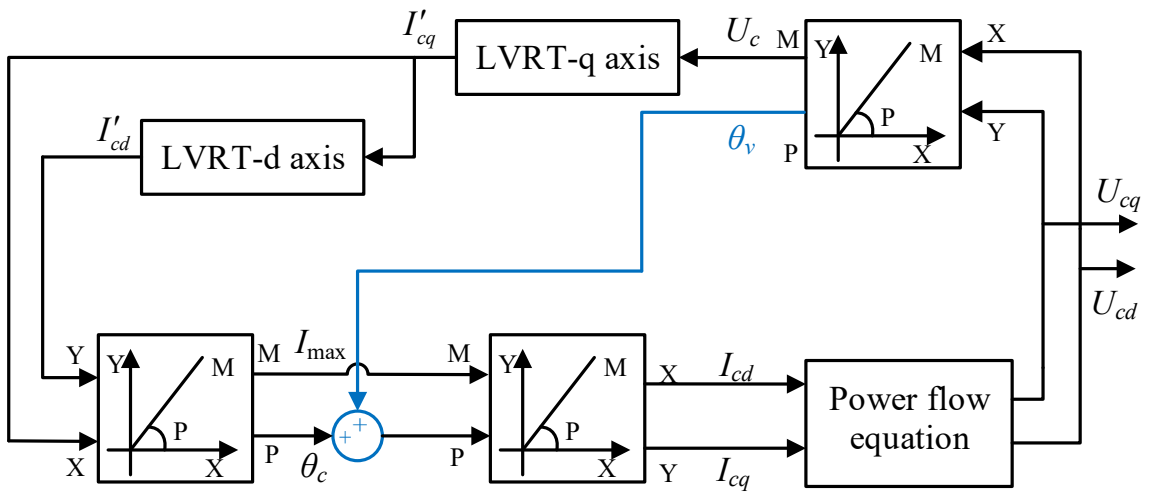

Fig. 10. Diagram of LVRT control.

The existence of EP during the on-fault period is first analyzed. By combining (8), (9), and (3), the on-fault power flow equation can be expressed by:

$$U_c e^{j(\theta_v+\delta_c)} = Z_{cf} I_{\max} e^{j(\delta_c+\theta_v+\theta_c+\theta_{zf})} \tag{11}$$

Where, $\theta_{zf}$ is the angle of the complex variable $\boldsymbol{Z_{cf}}$. From the first equation in (11), the solution of the equations exists only when $\theta_{zf}+\theta_c=0$. For example, when $\theta_{zf}$ is $\pi/2$, the solution exists only when $Z_{cf}I_{max}=U_c<U_{low}-I_{max}/K_q$ (i.e., under nearby fault), and when $\theta_{zf}<\pi/2$, the solution exists only under a specific on-fault $Z_{cf}$. $\theta_c=\arctan(I_{cq}/I_{cd})$, $I_{cq}$, and $I_{cd}$ are obtained from (2).

After fault clearance, by combining (8) (9) and (3), the on-fault power flow equation can be expressed by:

$$U_c e^{j(\theta_v+\delta_c)} = Z_{eq} I_{\max} e^{j(\delta_c+\theta_v+\theta_c+\theta_z)} + \boldsymbol{U'_g} \tag{12}$$

Where, $\theta_z$ is the angle of the complex variable $\boldsymbol{Z_{eq}}$. $\theta_c=\arctan(I_{cq}/I_{cd})$, $I_{cq}$, and $I_{cd}$ are obtained from (2). Compared to (11), the equivalent system voltage $\boldsymbol{U'_g}$ is involved in the first equation. It is difficult to provide explicit solvability conditions of (12), but the following analysis proves the existence of the solution.

Firstly, from Fig. 6, Fig. 7, and Fig. 8, $\theta_v$ can be both positive and negative as $\delta_c$ varies, and the variation of $\theta_v$ is continuous. So, there exists a $\delta_c=\delta_{c1}$, satisfying both $\theta_v=0$ and the power flow equation.

Secondly, considering that $\theta_v$ can mutate after fault clearance, if $\theta_v$ mutates to the value $\delta_{c1}-\delta_c$, the phase of $\boldsymbol{U_c}$ and $\boldsymbol{I_{cd}}$ will be $\delta_{c1}$, which is exactly equivalent to the above state when $\theta_v=0$ and $\delta_c=\delta_{c1}$ under traditional LVRT control. This indicates that there is at least one solution of (12): $\theta_v=\delta_{c1}-\delta_c$.

From the above analysis, the decoupled control cannot guarantee an EP only in the on-fault period. To address this issue, a delay activation can be added to the decoupled control part since the fault clearing time in a real power system is generally less than a fixed time, e.g. as 0.1s. The control diagram of decoupled LVRT control is shown in Fig. 11. Two switches are included in the control diagram. Switch 1 decides the activated type of LVRT, and Switch 2 decides whether LVRT is activated, and when the switch is on position B, LVRT is activated. When switch 2 is in position B, switch 1 is in position A, activating the traditional LVRT. After 0.1s, the fault is cleared, and switch 2 is in position B, activating the decoupled LVRT control.

The transient stability of the decoupled and traditional LVRT control is also analyzed. Due to the complexity of the IM model, stability domains are only given under 100% impedance load conditions. From (1) and (3), the dynamic equation of the system in Fig. 1 can be expressed as:

$$\begin{cases} \dfrac{d\delta_c}{dt} = \omega_c \\ \dfrac{d\omega_c}{dt} = K_i U_{cq} - K_p U'_g \omega_c \cos\left(\delta_g - \delta_c\right) \\ U_{cd} = U'_g\cos\left(\delta_g - \delta_c\right) + Z_{eq}I_{cd}\cos\theta_z - Z_{eq}I_{cq}\sin\theta_z \\ U_{cq} = U'_g\sin\left(\delta_g - \delta_c\right) + Z_{eq}I_{cd}\sin\theta_z + Z_{eq}I_{cq}\cos\theta_z \\ U_c = \sqrt{U_{cd}^2 + U_{cq}^2} \end{cases} \tag{13}$$

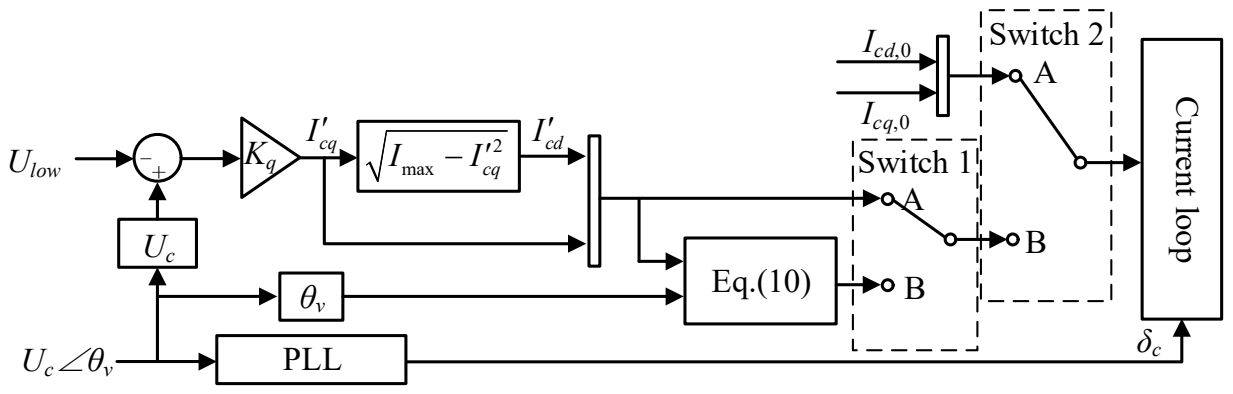

Fig. 11. Control diagram of decoupled LVRT control with delay activation.

By combining (13) with (2), (8), and (9) respectively, the transient models of the system in Fig. 1 under the traditional and decoupled LVRT controls can be obtained. The trajectory reverse method can be applied to give an accurate estimation of the stability domain [9], [21]. The stability domains of the system under the decoupled and traditional LVRT control are compared in Fig. 12. The circuit parameters are the same as the case in Fig. 6, and the PLL parameters $K_p$=20, $K_i$=200.

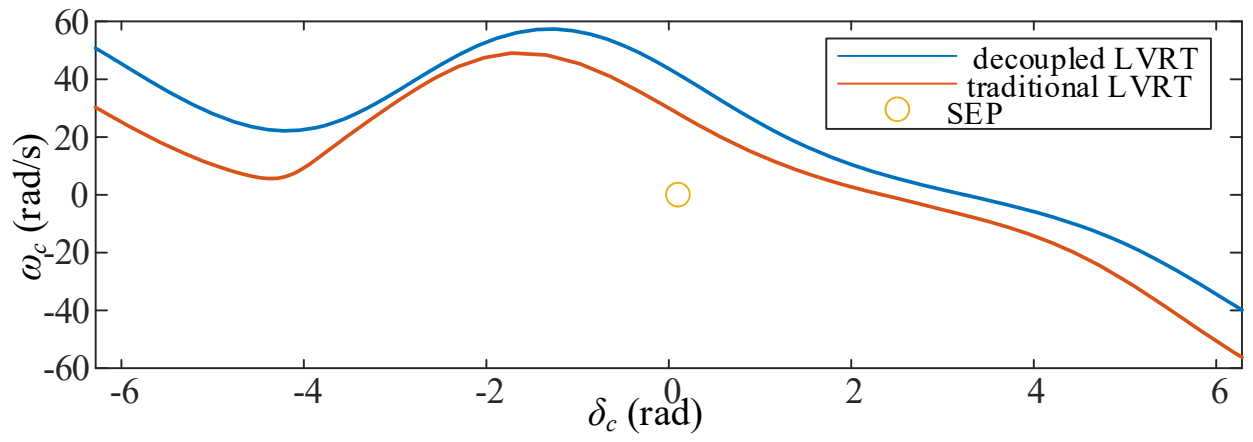

Fig. 12. Stability domains under the decoupled and traditional LVRT control

In Fig. 12, the area below the boundary is the stability domain, and it is obvious that under decoupled LVRT, the system has a larger stability domain. Considering that under decoupled control, the activated LVRT during the on-fault period is the traditional LVRT due to the delay activation, the fault trajectories of decoupled or traditional LVRT control are the same on the phase plane. This means the larger stability domain of decoupled LVRT corresponds to longer critical clearing time (CCT) and better transient stability.

## V. Case Study

An electromagnetic transient (EMT) simulation model of the system in Fig. 1 is developed in the PSCAD/EMTDC platform. The traditional and decoupled LVRT controls are tested in three cases, including 1) impedance load and nearby fault, 2) impedance load and remote fault, and 3) IM load and nearby fault. The fault location is the load node.

### *A. Case 1: impedance load and nearby fault*

The parameters of Case 1 are shown in Table 2. $S_c$ is the capacity of the converter. The reference current $I_{cd,ref}$ and $I_{cq,ref}$ correspond to 100 MW of active power output and 0 MVar of reactive power output. The load impedance $\boldsymbol{Z_l}$ is 0.495+j0.0495 p.u., which corresponds to 200MW of active load and 20MVar of reactive load. In Case 1, the impedance between the PCC point and the fault point $\boldsymbol{Z_{cf}}$ is $\boldsymbol{Z_c}$, which makes the on-fault $U_c$ lesser than $U_{low}$-$I_{max}/K_q$. The fault is classified as a nearby fault.

The fault simulation results of the decoupled and traditional LVRT controls are compared in Fig. 13, including the variation of $\delta_c$, $U_c$, $P_c$, and $Q_c$, and the fault duration is 0.15s. $\delta_c$ continuously decreases during the on-fault period, which aligns with the analysis results in Fig. 4 (c). After fault clearance, there is overvoltage under both control methods. This is because of the delay in the control loop, signal collection, and action of the converter, which prevents the converter from reducing its reactive power injection in time. However, under decoupled LVRT control, $U_c$ quickly decreases to the steady-state voltage, and the maximum $U_c$ during the transient process is 271.756 kV (1.182 p.u.). Under traditional LVRT control, the overvoltage time is longer, and the maximum $U_c$ reaches 286.656 kV (1.246 p.u.). The maximum $Q_c$ reaches 82.2425 MVar after fault clearance, and $Q_c$ does not decrease to zero until 0.075s after fault clearance. In contrast, the maximum $Q_c$ of decoupled control is 46.3407MVar, and $Q_c$ decreases to zero only 0.043s after fault clearance. In conclusion, simulation verifies the analysis that the PLL dynamic can leads to overvoltage problem under nearby fault, and the effect of the decoupled control is also verified.

TABLE II.
Parameters in Case 1 and Case 2

| Parameter | Value | Parameter | Value |
|---|---|---|---|
| $S_b$ | 100MVA | $S_c$ | 120MVA |
| $U_b$ | 230kV | $\omega_0$ | 120π rad/s |
| $\boldsymbol{U_g}$ | 1.03 p.u. | $K_q$ | 1.5 |
| $I_{max}$ | 1.0 p.u. | $I_{cd,ref}$ $I_{cq,ref}$ | 0.833p.u., 0p.u. |
| $K_p$ $K_i$ | Case1: 20, 1000<br>Case2: 20, 200 | $\boldsymbol{Z_c}$ $\boldsymbol{Z_g}$ | Case1: (0.1+j0.1)p.u., j0.1p.u.<br>Case2: j0.4p.u., j0.1p.u. |
| $\boldsymbol{Z_f}$ | 0Ω | $\boldsymbol{Z_l}$ | (0.495+j0.0495)p.u. |

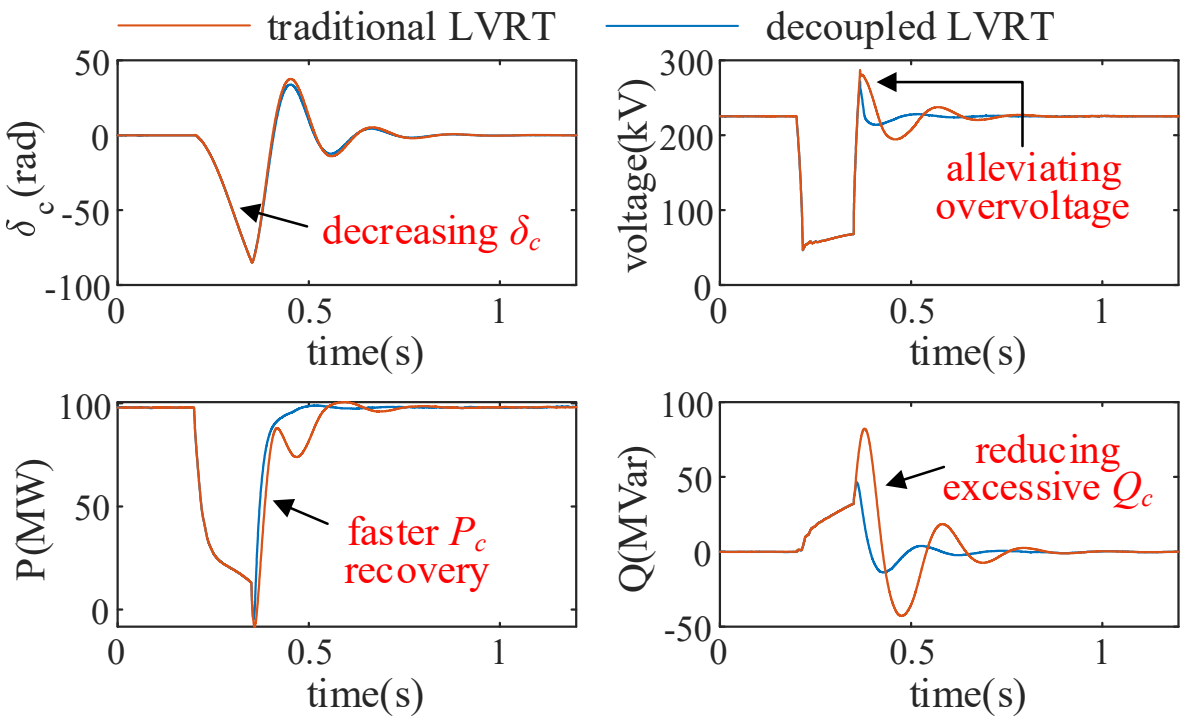

Fig. 13. Comparison of traditional and decoupled LVRT control in Case 1.

### *B. Case 2: impedance load and remote fault*

The parameters of Case 2 are shown in Table 2. $\boldsymbol{Z_c}$ increases to j0.4 p.u. to make the on-fault $U_c$ larger than $U_{low}$-$I_{max}/K_q$. The PLL parameter $K_i$ is also reduced to avoid loss of synchronization.

The fault simulation results of the decoupled and traditional LVRT controls are shown in Fig. 14, including the variation of $\delta_c$, $U_c$, $P_c$, and $Q_c$. The fault duration is 0.144s, which is exactly the CCT. $\delta_c$ continuously increases during the on-fault period, which aligns with the analysis in Fig. 4(d). After fault clearance, under traditional LVRT control, the voltage recovers from a low value, and both $P_c$ and $Q_c$ are negative at the beginning of the voltage recovery. Especially, the minimum $Q_c$ is lower than 50MVar, which means the reactive power absorption is close to the reactive power injection during the on-fault period. These problems align with the analysis in Fig. 6, and retard the voltage recovery. From Fig. 14, the voltage does not apparently recover until 0.2s after the fault clearance, which leads to a low-voltage problem. However, under the decoupled LVRT control, $U_c$ quickly recovers after a short time overvoltage caused by various delays in control and converter. In conclusion, the simulation verifies the analysis that the PLL dynamic will retard the post-fault voltage recovery under remote fault cases, and the effect of the decoupled control is also verified.

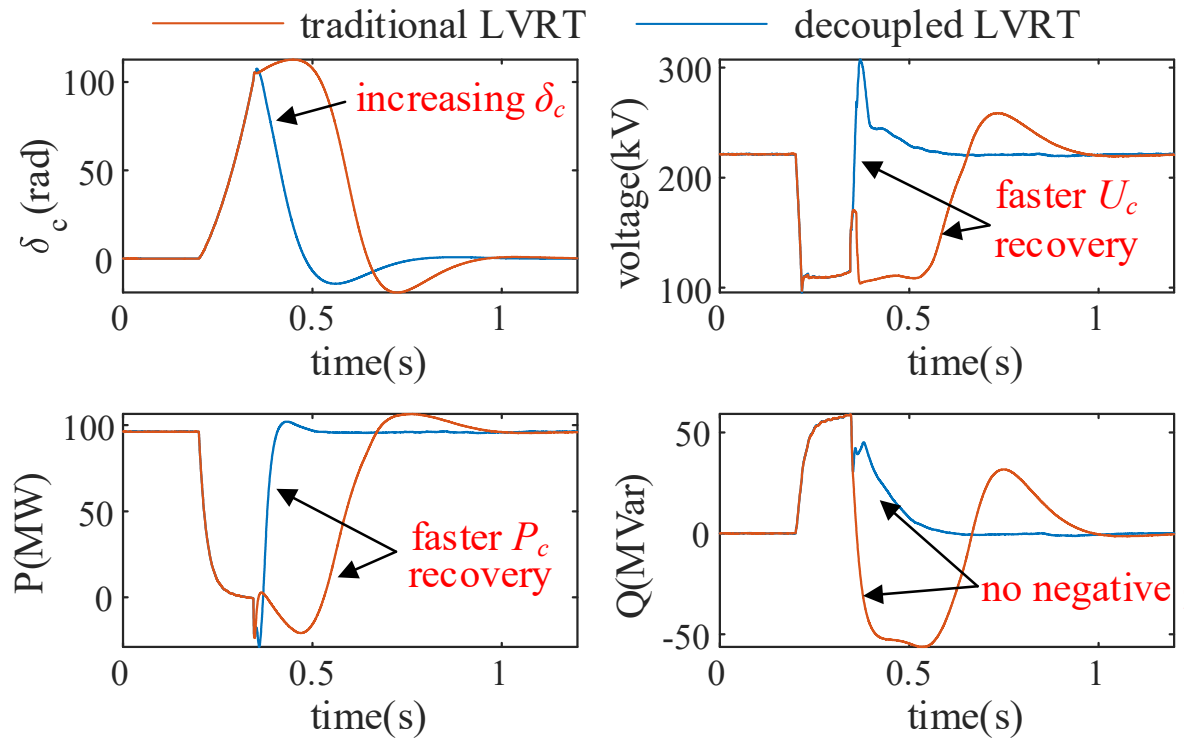

Fig. 14. Comparison of the traditional and decoupled LVRT control under 0.144s fault duration in Case 2

## C. Case 3: IM load and nearby fault

The parameters of Case 3 are shown in Table 3. The load is changed to IM load, and the system voltage is increased to meet the increased reactive power demand. The fault resistance is 10Ω to avoid instability in the IM.

TABLE III.
PARAMETERS IN CASE 3

| Parameter | Value | Parameter | Value |
|---|---|---|---|
| $S_b$ | 100MVA | $S_c$ | 120MVA |
| $U_b$ | 230kV | $\omega_0$ | 120π rad/s |
| $\boldsymbol{U_g}$ | 1.2 p.u. | $U_{dc}$ | 640kV |
| $K_q$ | 2 | $I_{max}$ | 1.0p.u. |
| $I_{cd,ref}$ $I_{cq,ref}$ | 0.833p.u., 0p.u. | $K_p$ $K_i$ | 20, 1000 |
| $\boldsymbol{Z_c}$ $\boldsymbol{Z_g}$ | j0.12p.u., j0.12p.u. | $\boldsymbol{Z_s}$ | j0.295p.u. |
| $X_m$ | 3.5p.u. | $X_r$ | 0.12p.u. |
| $R_r$ | 0.02p.u. | $\boldsymbol{Z_f}$ | 10Ω |

The fault simulation results of the decoupled and traditional LVRT controls are shown in Fig. 15, and DC voltage is also included in the last graph. The LVRT requirement is also added to the voltage graph. The fault durations are the maximum time for the two recovery processes to meet LVRT requirements, which are 0.180 s under traditional LVRT and 0.224s under decoupled LVRT, indicating a 24.444% increase of the maximum fault clearance time under the decoupled LVRT control. After fault clearance, $\delta_c$ is close to -180° under both control methods. Under traditional LVRT, $U_c$ recovers from a low value around 100kV. $P_c$ and $Q_c$ are negative at the beginning of the voltage recovery, and subsequent fluctuation of $Q_c$ results in a lower total reactive power injection compared to the expected value. All these problems retard the voltage recovery and align with the analysis in Fig. 8. Under decoupled LVRT, $U_c$ recovers from around 166kV, which is significantly larger than that under traditional LVRT. There are no prominent negative $P_c$ and low $Q_c$ problems. From the DC voltage comparison in the last graph, the negative $P_c$ under traditional LVRT leads to a delay in DC voltage reduction, and consequently, the maximum DC voltage is larger than that under the decoupled LVRT, even with a shorter fault duration. In conclusion, the simulation verifies the analysis that the PLL dynamic will retard the post-fault voltage recovery and increase the overvoltage risk at the DC side under IM proportion, large lagging $\delta_c$ cases. The effect of the decoupled control is proved.

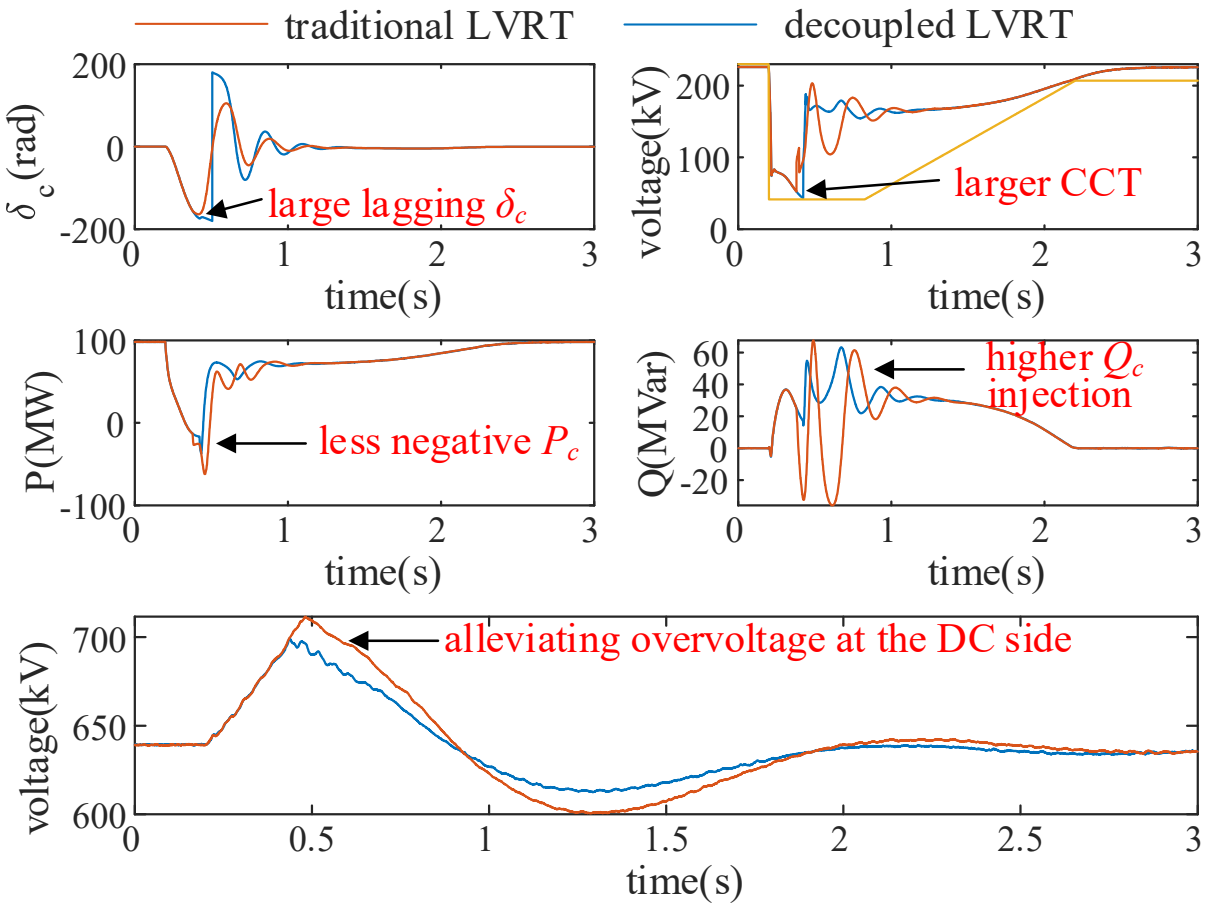

Fig. 15. Comparison of traditional and decoupled LVRT controls in Case 3.

## D. Case 4 and 5: remote and nearby fault in IEEE 9 bus

The proposed method is further tested in a modified IEEE 9 bus system. The topology of the system is shown in Fig. 16. A three-phase short circuit fault occurs in bus 7 in the simulation, and the fault duration is 0.1s. The synchronous generator (SG) on bus 2 is replaced by a GFL converter. The line impedance and load power can be found in [20], and the other parameters are shown in table IV. Besides, an impedance is added between bus 7 and bus 2. In Case 4, the impedance $Z_7$ is 0.03+j0.3 p.u. corresponding to remote fault cases. The load is composed of 50% impedance load and 50% IM load to make the low-voltage problem prominent. In Case 5, the impedance $Z_7$ in the modified IEEE 9 bus system is decreased to 0.05+j0.07 p.u., corresponding to the nearby fault condition. The load is composed of 100% impedance load to magnify the overvoltage problem.

In Case 4 and Case 5, the voltage curves under the traditional and decoupled LVRT controls are shown in Fig. 17 (a) and (b) respectively.

In Case 4, $\delta_c$ increases by 27° during the fault, and the voltage curves are similar to that in Case 2. From Fig. 17 (a), after fault clearance, a short-time overvoltage is observed due to the delay in the converter. Under traditional LVRT, due to the voltage angle coupling, the voltage drops sharply after the overvoltage and begins to recover from 155 kV (0.674 p.u.), which is close to the on-fault voltage. However, under the decoupled LVRT, the voltage only drops to 195 kV (0.848 p.u.) before the recovery, which is significantly higher compared to the traditional LVRT case and makes the low-voltage problem less prominent.

In Case 5, the voltage curves are similar to those in case 1. In Fig. 17 (b), overvoltage is observed due to the delay in the converter. However, under traditional LVRT, the voltage-angle coupling further increases the voltage to 352 kV (1.530

p.u.), which obviously exceeds general voltage stability codes. Under the decoupled LVRT control, the overvoltage problem is significantly alleviated, and the maximum voltage is reduced to 281 kV (1.222 p.u.) compared to the traditional LVRT case.

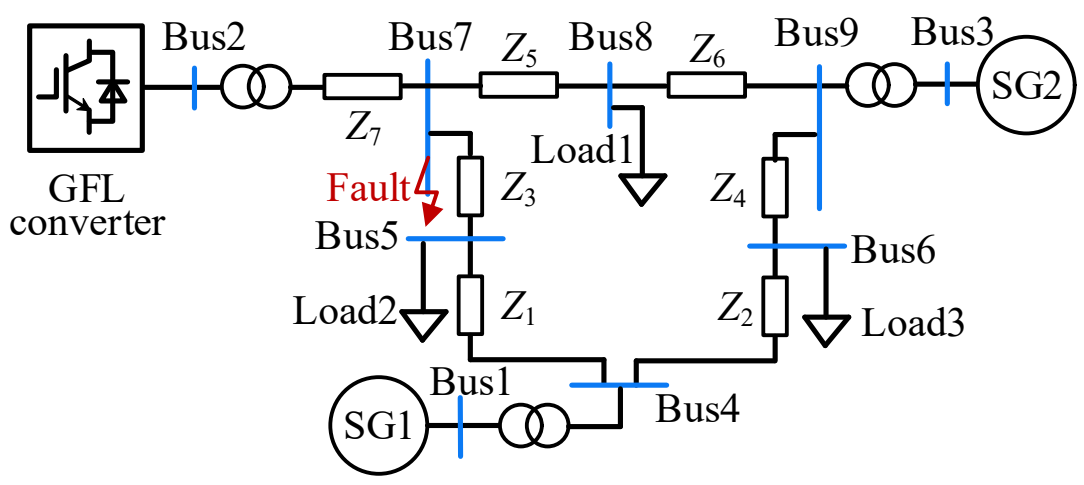

Fig. 16. Topology of the modified IEEE 9 bus system.

TABLE IV.
PARAMETERS OF THE MODIFIED IEEE 9 BUS TEST SYSTEM

| Parameter | Value | Parameter | Value |
|---|---|---|---|
| $S_b$ | 100MVA | $S_c$ | 200MVA |
| $U_b$ | 230kV | $\omega_0$ | 100π rad/s |
| $Z_7$ | 0.03+j0.3p.u. | $R_f$ | 1Ω |
| $K_q$ | 1.5 | $I_{max}$ | 1.2p.u. |
| $I_{cd,ref}$ $I_{cq,ref}$ | 0.817p.u., 0.123rad | $K_p$ $K_i$ | 20, 500 |
| $X_m$ | 3.5p.u. | $\boldsymbol{Z_s}$ | j0.295p.u. |
| $R_r$ | 0.02p.u. | $X_r$ | 0.12p.u. |

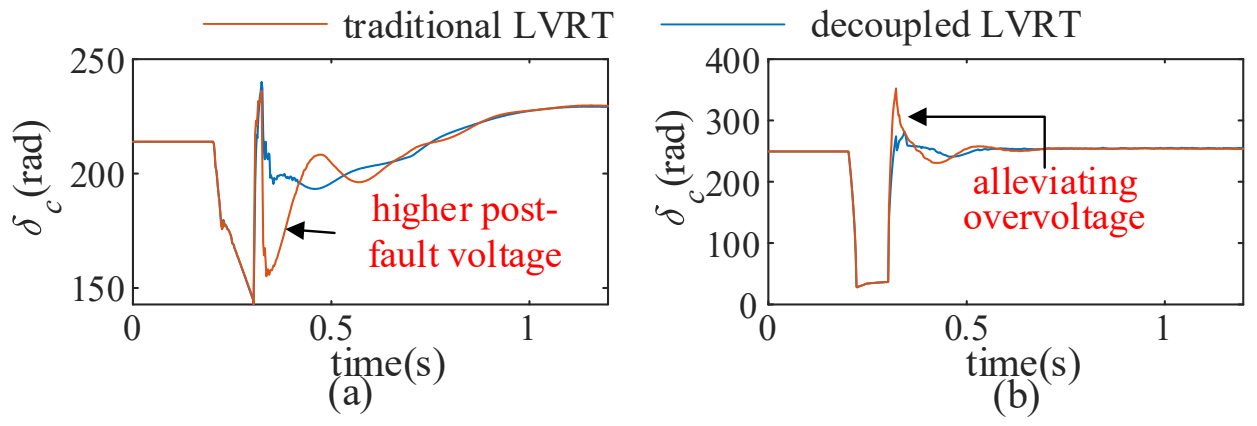

Fig. 17. Voltage curve under traditional and decoupled LVRT in case 4

## VI. CONCLUSION

The voltage angle coupling mechanism of the GFL reources caused by PLL dynamic and LVRT control is revealed in this paper, and a novel decoupled control is proposed to eliminate the influence of voltage angle coupling. The following conclusions can be drawn:

1) The transient voltage tendency under different fault types is analyzed. More specifically, there is a low voltage tendency under remote fault, and a high voltage tendency may occur under nearby fault.

2) The influence of IM dynamic on voltage angle coupling is analyzed. The IM dynamic can narrow the range of the PLL phase, causing high voltage tendency, and low voltage tendency may also occur under severe nearby fault due to IM dynamic.

3) A power decoupled control is proposed to eliminate the influence of the voltage angle coupling. The existence of the equilibrium point and transient stability domain is also analyzed. The effect of the decoupled control method is verified in case studies in both low-voltage and high-voltage scenarios.